\documentclass[usenatbib,referee]{mn2e}
\usepackage{graphicx}
\usepackage{epsfig}
\usepackage{color}

\title[Evolving ONe WD+He WD systems to UCXBs]
{Evolving ONe WD+He WD systems to ultra-compact X-ray binaries}

\author[D. Liu \& B. Wang]
{D. Liu$^{\rm 1,2}$\thanks{E-mail:liudongdong@ynao.ac.cn}, B. Wang$^{\rm 1,2}$\thanks{E-mail:wangbo@ynao.ac.cn}\\
$^1$Yunnan Observatories,  Chinese Academy of Sciences, Kunming 650216, China\\
$^2$Key Laboratory for the Structure and Evolution of Celestial Objects, Chinese Academy of Sciences, Kunming 650216, China}

\begin{document}
\date{}
\pagerange{\pageref{firstpage}--\pageref{lastpage}} \pubyear{2022}
\maketitle

\label{firstpage}

\begin{abstract}\label{0. abstract}
It has been proposed that accretion-induced collapse (AIC) of massive white-dwarfs (WDs) is an indispensable path for the formation of neutron star (NS) binaries.
Although there are still no direct evidence for the existence of AIC events, several kinds of NS systems are suggested to originate from the AIC processes. 
One of the representative evidence is the detection of the strong magnetic field and slow spin NSs with ultra-light companions ($\le 0.1\,\rm M_{\odot}$) in close orbits.
However, previous studies cannot explain the formation of AIC events with such low-mass companions.
In the present work, we evolve a series of ONe WD$+$He WD systems to the formation of AIC events (named as the He WD donor channel), and the NS binaries behave as ultra-compact X-ray binaries when the He WDs refill their Roche-lobes.
We found that the ONe WD$+$He WD systems a possible channel for the formation of the newly formed NS$+$ultra-light companion systems just after the AIC event.
Although there are some other inconsistent properties, the detected companion mass and orbital period of 4U 1626-67 (one of the newly formed NS binaries with ultra-light companions) can be reproduced by the He WD donor channel.
In addition, combined with previous asteroseismology results, we speculate that an UCXB source (XTE J1751$-$305) may originate from the He WD donor channel.


\end{abstract}


\section{Introduction} \label{1. Introduction}
The accretion-induced collapse (AIC) events are a type of electron capture supernovae resulting in the formation of neutron star (NS) systems (e.g. Nomoto et al. 1979; Miyaji et al. 1980; Taam \& van den Heuvel 1986),
which occurs when the core density of ONe WDs with masses approach to the Chandrasekhar mass limit becomes high enough for the occurrence of electron captures onto Ne and Mg, taking away the pressure support and leading to the collapse of ONe WDs into NSs (e.g. Nomoto \& Kondo 1991; Ivanova \& Taam 2004).
Compared with typical supernovae, the AIC events may eject much less material ($<0.1\,M_{\odot}$), synthesizing quite small $^{\rm 56}$Ni ($<0.01\,M_{\odot}$) moving at very high velocity ($\sim$0.1 light velocity), which indicates that the optical transients of AIC events would be much fainter than a typical supernova, and the duration would be much shorter (e.g. Metzger, Piro \& Quataert 2009; Darbha et al. 2010; Piro \& Kulkarni 2013).
Up to now, there are still no direct observational evidence to support the existence of AIC events.
The existence of AIC events is supported by some indirect evidences, as follows (for a recent review see Wang \& Liu 2020).

Some population synthesis studies indicate that most of the retained NSs in the globular clusters must originate from the electron-capture supernovae (e.g. Ivanova et al. 2006, 2008; Kuranov \& Postnov 2006). Meanwhile, there exist some peculiar pulsars with low space velocities or with relatively young ages, which are strong candidates of post-AIC events.
(1) Traditional iron core-collapse supernovae (CCSNe) have large kick velocities, meaning that NSs from the CCSN channel are more likely to be kicked out from the globular clusters or have large space velocities if they remain in the globular clusters (e.g. Tauris \& Takens 1998).
A large number of NSs have been detected in the globular clusters, especially some of them are recycled pulsars with low space velocities, which can be naturally explained by the AIC events owing to their small momentum kick and little mass loss (Bailyn \& Grindlay 1990; Kitaura, Janka, \& Hillebrandt 2006; Dessart et al. 2006).
(2) The progenitor lifetimes of CCSNe are significantly shorter than the age of old globular clusters, which means that the NSs from the CCSN channel have similar ages to their host globular clusters.
Therefore, the detection of obviously young NSs in some globular clusters is a strong evidence for the AIC events as the progenitors of AIC events have significant longer lifetimes (Boyles et al. 2011; Wang 2018).

In the Galactic disk, some particular NS binaries are also proposed as candidates of AIC productions.
These NSs with slow spin and relatively strong magnetic fields have ultra-light companions ($\le\,0.1\,\rm M_{\odot}$) and close orbits, like 4U 1626-67 (Taam \& van den Heuvel 1986), GRO J1744-28 (van Paradijs et al. 1997), PSR J1744-3922 (Breton et al. 2007) and PSR B1831-00 (Sutantyo \& Li 2000), etc.
On the one hand, the ultra-light companions and close orbits indicate that a considerable dynamical stable mass-transfer process has taken place between these binaries (as a dynamical unstable mass-transfer are more likely lead to a merger during the CE process for such small mass-ratio binaries).
On the other hand, the slow spin and relatively strong magnetic fields of these NSs both imply that the NSs are newly burned without accreting large amount of material\footnote{Such long spin period are reminiscent of the observed slow spinning magnetars (e.g. Mereghetti \& Stella 1995), which may provides another alternative approach to explain the formation of these slow spin and relatively strong magnetic field NS systems.}.
Although the AIC events are usually thought to produce NSs with fast spin and unclear magnetic field, it has been suggested that the formation of these newly burn NSs with ultra-light companions might be explained by the AIC events with surviving donors, in which the considerable dynamical stable mass-transfer process occurs in the WD binary stage, and the newly formed NSs accompanying with ultra-light stars are detected just after the AIC process and before (or at the beginning of) the recycling process (e.g. Tauris et al. 2013).


There are two popular progenitor models for the production of AIC events, i.e. the double-degenerate model and the single-degenerate model.
In the double-degenerate model, the merging of ONe WD$+$ONe/CO WD (and even double CO WDs) with super-Chandrasekhar masses are supposed to produce AIC event, which would lead to the formation of isolated NSs with special properties (e.g. Ruiter et al. 2019; Liu \& Wang 2020).
In the single-degenerate model, an ONe WD increases its mass by accreting material from its non-degenerate companion, which could be a main sequence (MS), a subgiant, a red-giant (RG) or a helium (He) star. 
The WD would collapse into an NS when its mass approach to the Chandrasekhar limit (e.g. Nomoto \& Kondo 1991; Ivanova \& Taam 2004).
Tauris et al. (2013) suggested that the ONe WD$+$MS systems would produce fully recycled MSP$+$He WD systems through the AIC process, while the ONe WD$+$RG/He star systems may evolve to mildly recycled pulsars with CO/ONe WD companions.
Ablimit \& Li (2015) found that the ONe WD$+$MS channel of the AIC events can account for LMXBs with strong-field NSs and MSP$+$He WD systems in short orbits after considering the excited wind from the MS induced by X-ray irradiation from the accreting WD.
Wang, Liu \& Chen (2022) suggested that the ONe WD$+$RG channel could contribute to the observed MSPs with orbital periods ranging from 50 to 1200\,d.
Liu et al. (2018) found that the ONe WD$+$He star channel could explain the formation of intermediate-mass binary pulsars with short orbital periods, and can account for the observed parameters of PSR J1802$-$2124 (one of the two well$-$observed intermediate$-$mass binary pulsars).

However, previous studies on the progenitors of the single-degenerate model of the AIC events predicted extremely more massive companions after the AIC process (Tauris et al. 2013; Wang 2018).
The companion star from the ONe WD$+$MS channel would have masses in the range of 1.3$-$$2.8\,\rm M_{\odot}$ after the AIC process (Tauris et al. 2013; Ablimit \& Li 2015), and the ONe WD$+$RG channel would lead to the formation of 0.8$-$$1.2\,\rm M_{\odot}$ companions (Wang, Liu \& Chen 2022), and the ONe WD$+$He star channel corresponds to 0.8$-$$2.3\,\rm M_{\odot}$ companions (Liu et al. 2018).
In this case, the formation of these newly formed NS with ultra-light companions becomes a challenge to our understanding of close binary evolution, as discussed in Tauris et al. (2013).

It is worth noting that the mass-transfer from a He WD onto a ONe WD is always dynamically stable (see Marsh, Nelemans \& Steeghs 2004).
If the ONe WD$+$He WD systems would evolve to AIC events, the He WDs would also survive and may refill their Roche-lobe.
In the present work, we evolve a series of ONe WD$+$He WD systems to the formation of AIC events, and found that this channel can evolve to NS$+$ultra-light companion systems just after the AIC process.
In Sect.\,2, we present the numerical methods used in the present work.
The corresponding results are given in Sect.\,3.
A discussion and a summary are given in Sect.\,4.

\section{Numerical code and methods} \label{2. Methods}
We employ the stellar evolution code Modules for Experiments in Stellar Astrophysics (MESA) to simulate the evolution of ONe WD$+$He WD systems (version 12778; Paxton et al. 2011, 2015, 2019). We use the test suite make\_he\_wd to build He WD models with metallicity $Z=0.02$. In the present work, we consider the loss of the orbital angular momentum caused by the gravitational wave radiation (see Landau \& Lifshitz 1971) and mass loss.

We assume that the mass-transfer process of ONe WD$+$He WD systems and NS$+$He WD systems are dynamically stable (see Marsh, Nelemans \& Steeghs 2004; Chen et al. 2022), and use the Ritter scheme to calculate the mass-transfer from He WDs onto ONe WDs/NSs.
The accreted He-rich material will burn stably on the surface of ONe WDs, resulting in the mass-growth of the ONe WDs.
Similar to our previous study (see Liu et al. 2018), the prescription for the mass-growth of ONe WDs is from Nomoto (1982) and Kato \& Hachisu (2004).
An AIC event is supposed to occur when the ONe WD grows in mass approach to the Chandrasekhar mass limit, which is set to be  $1.378\,\rm M_{\odot}$ (e.g. Wu \& Wang 2018).

We assume that the Chandrasekhar mass WD will collapse into a $1.25\,\rm M_{\odot}$ NS during the AIC process (e.g. Ablimit \& Li 2015).
After the AIC process, we simply assume that the orbit remains circular,
and adopt a specific angular momentum conservation prescription presented in Verbunt, Wijers \& Burn (1990) to calculate the orbit change, written as
\begin{equation}
\frac{a}{a_{\rm 0}}=\frac{M_{\rm WD}+M_{\rm 2}}{M_{\rm NS}+M_{\rm 2}},
\end{equation}
where the $a_{\rm 0}$, $a$, $M_{\rm WD}$, $M_{\rm 2}$ and $M_{\rm NS}$ are the orbital separation before and after the AIC process, the mass of the ONe WD, the He WD and the newly formed NS star at the moment of just before and after AIC, respectively.
Note that this equation is a non-standard model by assuming that the orbit remains circular.
In reality, the AIC occurs essentially instantaneously as its timescale is significantly shorter than the orbital period.
Thus, the orbit changes in the same way as a sudden mass-loss induced supernova kick, and the instantaneous separation and relative velocity remain constant.
For more details for investigating the orbital changes see, e.g. Brandt et al. (1995), Tauris \& Takens (1998).
However, this does not make a big difference as shown in Tauris et al. (2013), and the influence of kick velocity is not considered in the present work.

According to Eq.\,(1), the orbital separation becomes longer after the AIC process, leading to the consequence that the He WD turns to below its Roche-lobe.
Subsequently, the orbital separation decreases due to the gravitational wave radiation.
The He WD would fill its Roche-lobe again and transfer He-rich material onto the NS.
During this phase, the binary behaves as an ultra-compact X-ray binary (UCXB; e.g. Tutukov et al. 1985; Savonije, et al. 1985; Nelson et al. 1986; Nelemans \& Jonker 2010; Sengar et al. 2017), which is defined as the low-mass X-ray binaries with orbital period less than $\sim$1\,h and is potential target for space gravitational wave detectors (e.g. Chen, Liu \& Wang 2020).
We assumed that a fraction $\beta=0.5$ of the transferred material could be accreted onto the NS and that the accretion rate is Eddington limited (e.g. Podsiadlowski, Rappaport \& Pfahl 2002; Tauris \& van den Heuvel 2006).
The mass-growth rate of the NS ($\dot{M}_{\rm NS}$), written as
\begin{equation}
\dot{M}_{\rm NS}=(|\dot{M}_{\rm 2}|-\max[|\dot{M}_{\rm 2}|-\dot{M}_{\rm Edd},0])\cdot \beta,
\end{equation}
in which $\dot{M}_{\rm Edd}$ is the Eddington mass-accretion rate for He-accreting NSs (e.g. Tauris et al. 2013), written as
\begin{equation}
\dot{M}_{\rm Edd}=4.6\times 10^{\rm -8}\,M_{\odot}\,\rm yr^{\rm -1}\cdot (M_{\rm NS}/M_{\odot})^{\rm -1/3}.
\end{equation}
Note that this equation is obtained based on a simply assumption that the NS mass-radius relation is a perfect n=1.5 polytrope, which is a rough estimate.


\section{Results} \label{3. Results}
\begin{table*}
\begin{center}
\caption{Selected Evolutionary Properties of  Some Binaries with Different Initial Primary/Secondary Masses and Orbital Periods.
Set 5 represents a possible case for the formation of 4U 1626$-$67.
The columns (from left to right):  the initial ONe WD mass, the initial donor mass and the initial orbital period;
the mass-transfer time-scale of ONe WD$+$He WD systems;
the donor mass and orbital period just after AIC process;
the mass-transfer time-scale of NS$+$He WD systems;
the final mass of NS and donor star, and the final orbital period.}
\begin{tabular}{ccccccccccccc}
\hline\hline
Set & $M_{\rm ONe\,WD}^{\rm i}$ & $M_2^{\rm i}$ & $P_{\rm orb}^{\rm i}$ & $\delta t_{1}$ & $M_2^{\rm AIC}$ & $P_{\rm orb}^{\rm AIC}$ & $\delta t_{2}$ & $M_{\rm NS}^{\rm f}$ & $M_2^{\rm f}$ & $P_{\rm orb}^{\rm f}$\\
 &  ($\rm M_{\odot}$)   &  ($\rm M_{\odot}$)   &  (d)  & (Myr)  &  ($\rm M_{\odot}$)   &  (d)  & (Gyr)   &  ($\rm M_{\odot}$)   &  ($\rm M_{\odot}$)   &  (d)\\
\hline
1  & 1.30  & 0.20  & 0.015 & 0.21                  & 0.1172       &0.0051  &1.65                          & 1.3033    &   0.0082  & 0.0385\\
2  & 1.30  & 0.18 & 0.030 & 0.92                   &0.0819        &0.0084   &5.08                       & 1.2880    &    0.0059 & 0.0481\\
3  & 1.34  & 0.18 & 0.080 & 0.11             &0.1399        &0.0046   &4.13                       &  1.3110     &     0.0062 & 0.0468\\
4  & 1.33  & 0.15 & 0.015 & 0.41             &0.0933      &0.0066   &3.85                       &  1.2935      &   0.0063  & 0.0461\\
5  & 1.26562 & 0.20 & 0.045 & 407.9           &0.0127      &0.0339  &1.15 &1.2522 & 0.0084 & 0.0380\\
\hline
\end{tabular}
\end{center}
\end{table*}

\subsection{Binary evolution results}
\begin{figure*}
\includegraphics[width=12.5cm,angle=0]{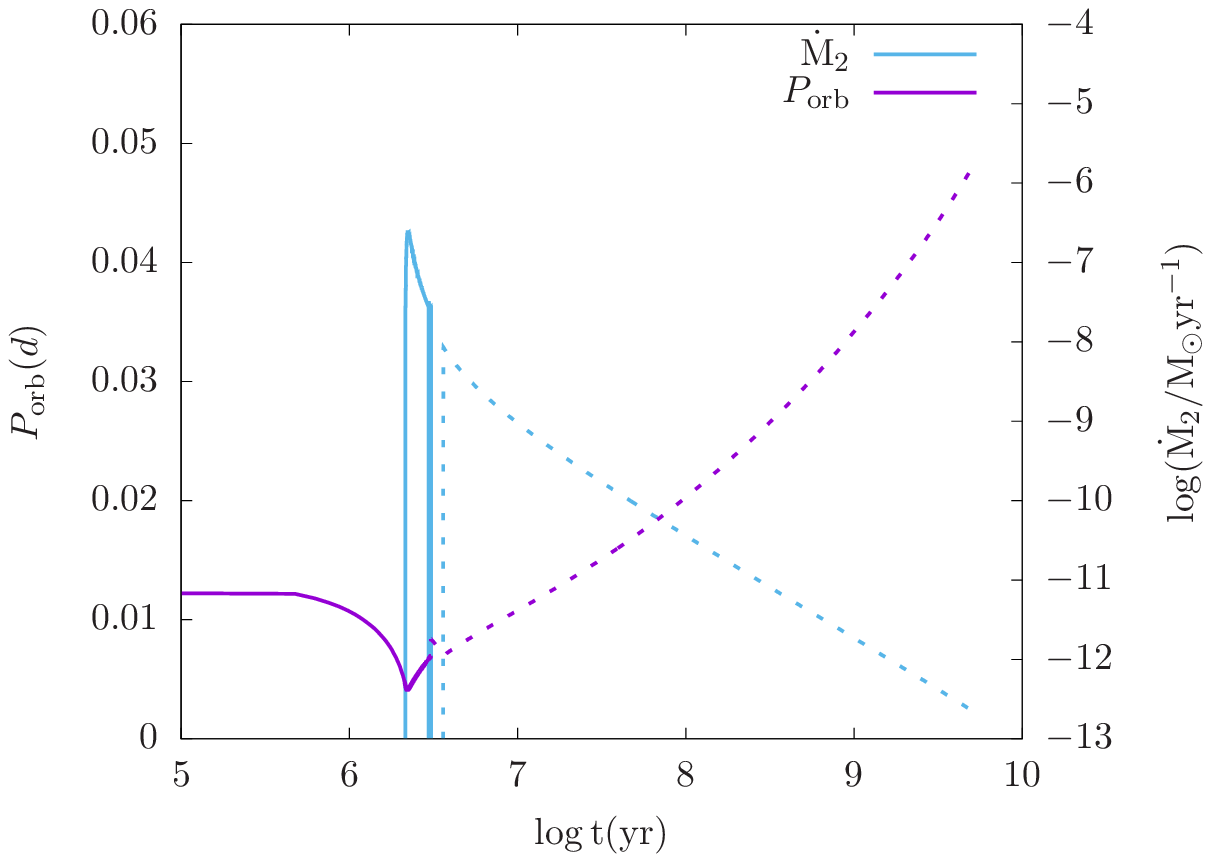}
\includegraphics[width=12.5cm,angle=0]{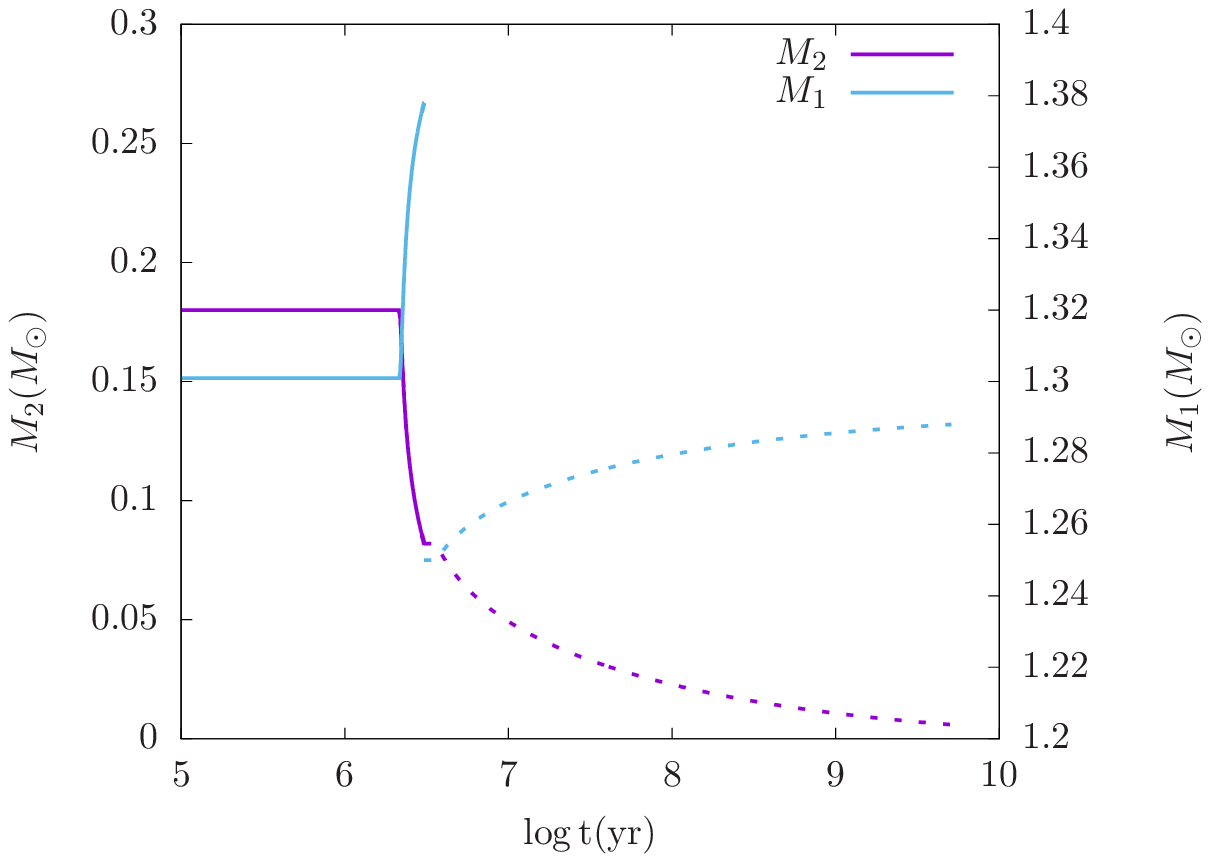}
\caption{A typical example for the evolution of an ONe WD$+$He WD system to a UCXB via the AIC process, in which the initial parameters ($M_{\rm ONe WD}^{\rm i}$, $M_2^{\rm i}$, $P_{\rm orb}^{\rm i}$)=(1.3$\,\rm M_{\odot}$, 0.18$\,\rm M_{\odot}$, 0.03\,d) (see Set.\,2 in Tab.\,1).
The upper panel presents the evolution of the mass-transfer rate ($\dot{M}_{\rm 2}$) and orbital period ($P_{\rm orb}$) as a function of time.
The lower panel shows the evolution of the primary star mass ($M_{\rm P}$) and the He WD mass ($M_{\rm He\,WD}$) as a function of time. The solid and dotted lines represent evolutionary phases before and after the AIC process, respectively.}
\end{figure*}

We conducted a series of complete binary evolution for ONe WD+He WD systems to the formation of UCXBs via the AIC processes.
Tab.\,1 lists the main evolutionary properties of five typical ONe WD+He WD systems that can evolve into UCXBs.
In Fig.\,1, we present a typical example for the binary evolution of a 1.3$\,\rm M_{\odot}$ ONe WD and a 0.18$\,\rm M_{\odot}$ He WD with an orbital period of 0.03\,d (see Set.\,2 in Tab.\,1).
At first, the binary separation gradually decrease because of the gravitational wave radiation.
At about t=3.13 Myr, the He WD begins to fill its Roche-lobe, and transfer He-rich material onto the surface of the ONe WD, leading to the mass-growth of the ONe WD.
At about t=3.21 Myr, the binary evolves to have a minimum orbital period of $\sim$0.004 d. 
At this stage, the masses of the ONe WD and He WD are 1.3083$\,\rm M_{\odot}$ and 0.1693$\,\rm M_{\odot}$, respectively.
The ONe WD grows in mass to the Chandrasekhar mass limit at about t=4.05 Myr.
Subsequently, an AIC event is assumed to occur and the ONe WD would collapse to a 1.25$\,\rm M_{\odot}$ NS with a 0.0819$\,\rm M_{\odot}$ He WD companion.

After the AIC process, the orbital period becomes larger as a result of the mass reduction.
At about t=4.61 Myr, the He WD refills its Roche-lobe due to the orbital shrinking driven by the gravitational wave radiation.
During the following 5.08 Gyr, the system behaves as an UCXB.
Finally, the binary have a 1.2880$\,\rm M_{\odot}$ NS and a 0.0059$\,\rm M_{\odot}$ donor with an orbital period of 0.0481 d.
The evolutionary code stops at this stage because of hitting the equation-of-state limits.
Note that Go, Wang \& Han (2022) suggested that some black windows with donor masses less than 0.01$\,\rm M_{\odot}$ could be produced from the UCXBs with He star companions. 
In this case, we can speculate that the UCXBs from the AIC channel would also evolve to black windows in their following evolutions if the evaporation process is considered.
It can be expected that the binary will eventually evolve to a single millisecond pulsar (Ruderman \& Shaham 1985) or a pulsar$+$planet$-$like system (e.g. Podsiadlowski 1993).

\begin{figure}
\begin{center}
\epsfig{file=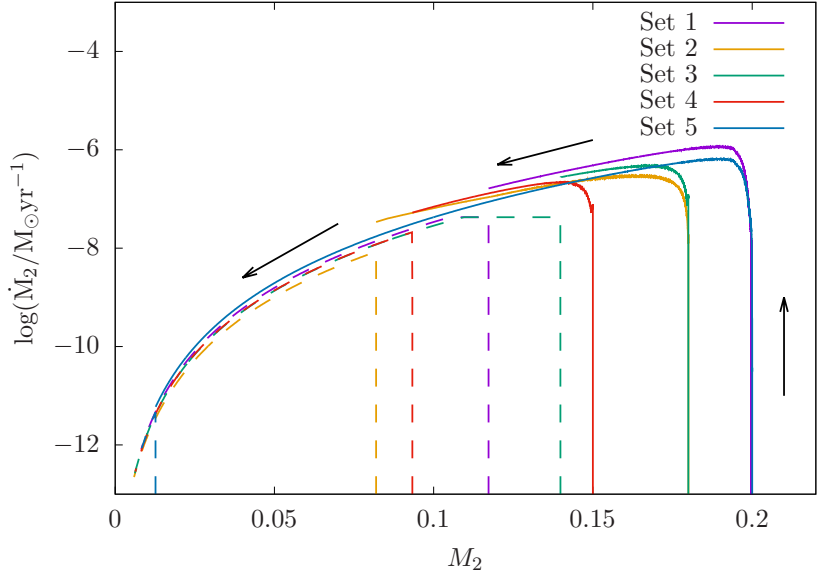,angle=0,width=12.5cm}
 \caption{The evolution of the five typical ONe WD$+$He WD systems that can produce UCXBs via the AIC process in the He WD mass and the mass-transfer rate ($M_{\rm He\,WD}$$-$$\dot M_{\rm 2}$) plane. The solid and dashed lines represent evolutionary phases before and after the AIC process, respectively.}
\end{center}
\end{figure}

Fig.\,2 presents the binary evolution of five typical ONe WD$+$He WD systems that can produce AIC events in the He WD mass and the mass-transfer rate plane.
The details of their evolutionary properties are shown in Tab.\,1.
It is notable that the mass-transfer rate have strong correlation with the He WD mass, which would be useful in the observations to determine the He WD mass.
From this figure, we can see that the mass-transfer rate generally decreases with the He WD mass after the very beginning of the mass-transfer process, which is caused by the increase of the orbital period (see Sengar et al. 2017).
As shown in Tab.\,1, the He WD mass at the AIC moment ($M_2^{\rm AIC}$) ranges from $\sim$0.01$\,\rm M_{\odot}$ to $\sim$0.15$\,\rm M_{\odot}$. 
Thus, the He WD donor channel of the AIC events is an alternative path for the formation of the newly formed NSs with ultra-light companions.

Here, sets 1 and 2 show the cases with the same ONe WD mass but different initial He WD masses.
Comparing these two cases, we found that as the initial He WD mass decreases, $M_2^{\rm AIC}$ and the final NS mass ($M_{\rm NS}^{\rm f}$) both decreases, while the mass-transfer timescale between ONe WD$+$He WD systems ($\delta t_{1}$) and between NS$+$He WD systems ($\delta t_{2}$) both increases.
Set 2 and set 3 represent the cases with same He WD mass but different initial ONe WD masses, which indicates that $M_2^{\rm AIC}$ and $M_{\rm NS}^{\rm f}$ increase, whereas $\delta t_{1}$ and $\delta t_{2}$ decrease when the initial ONe WD mass increases.
Set 4 represents the case with a relative low-mass He WD (0.15$\,\rm M_{\odot}$), in which the initial ONe WD mass should be as massive as $\sim$1.33$\,\rm M_{\odot}$.
Set 5 displays a possible case for the formation of 4U 1626$-$67, which will be discussed in Sect.\,3.2.

\begin{figure}
\begin{center}
\epsfig{file=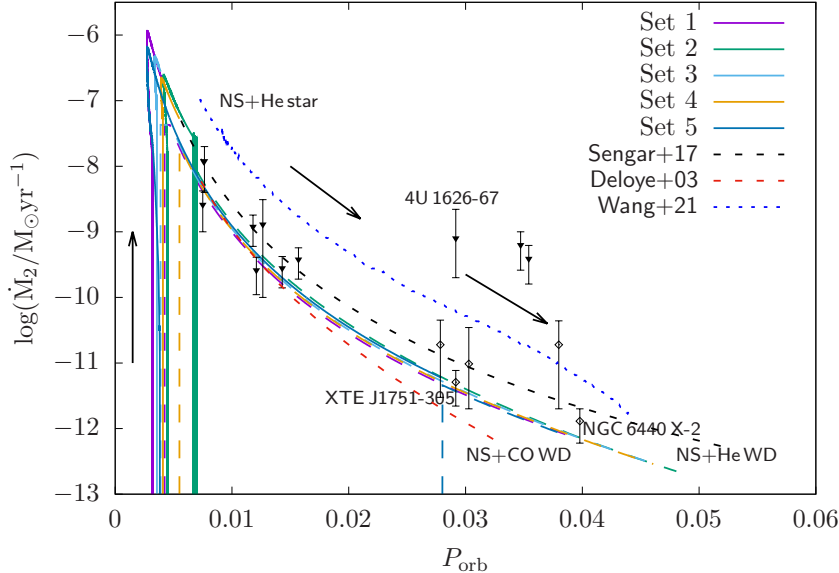,angle=0,width=12.5cm}
 \caption{The evolution of the five typical ONe WD$+$He WD systems evolving to UCXBs in the orbital period$-$the mass-transfer rate ($P_{\rm orb}$$-$$\dot M_{\rm 2}$) plane. The solid and dotted lines display evolutionary phases before and after the AIC process, respectively.
We also present the evolutionary tracks for the formation of UCXBs from the NS$+$CO WD channel (red dashed line; Deloye \& Bildsten 2003), NS$+$He WD channel (black dashed line; Sengar et al. 2017) and NS$+$He star channel (blue dotted line; Wang et al. 2021).
Meanwhile, the observed UCXBs presented in Heinke et al. (2013) are also plotted here; the filled triangles and open triangles represent  the persistent sources and the transient sources, respectively.}
  \end{center}
\end{figure}

Fig.\,3 displays the evolution of ONe WD$+$He WD systems to UCXBs via the AIC channel in the orbital period$-$the mass-transfer rate ($P_{\rm orb}$$-$$\dot M_{\rm 2}$) plane.
The evolutionary tracks of UCXBs originated from the AIC channel locate between those from the NS$+$He WD channel and the NS$+$CO WD channel.
For the evolutionary phase before the AIC process, the tracks are consistent with that from the NS$+$He WD channel.
However, the effective temperature and the surface degeneracy of He WDs both decrease during the period from the AIC process to the moment when the He WDs refill their Roche-lobe, leading to a different evolutionary track during the UCXB phase.
From this figure, we can see that the evolution tracks from our simulation can cover the location of about 10 observed UCXBs, in which XTE J1751$-$305 cannot be reproduced by other formation channels.
More discussions about XTE J1751$-$305 see Sect.\,4.

\begin{figure}
\begin{center}
\epsfig{file=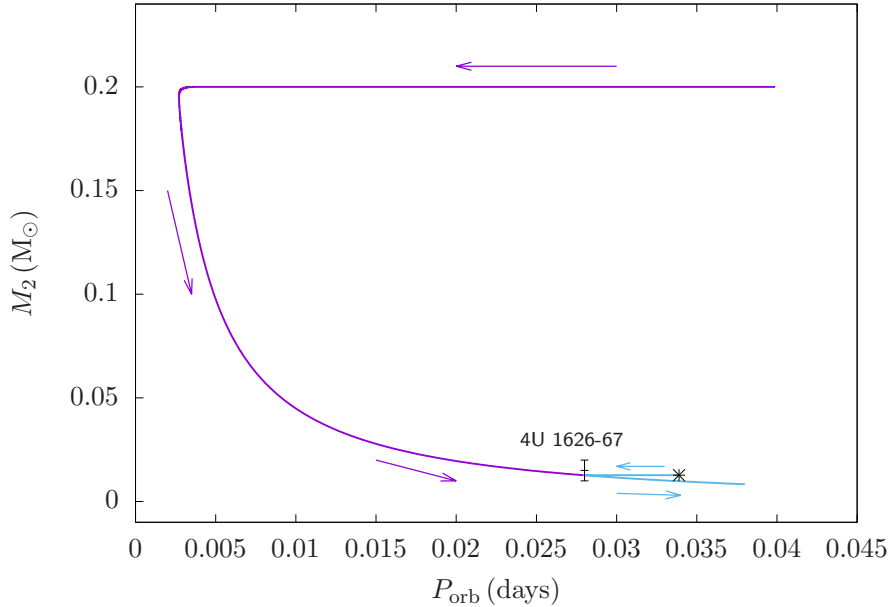,angle=0,width=12.5cm}
 \caption{A possible evolutionary track for the formation and future evolution of 4U 1626$-$67.
The snowflake symbol shows the start position of the evolutionary track just after the AIC process.
The initial parameters ($M_{\rm ONe\,WD}^{\rm i}$, $M_2^{\rm i}$, $P_{\rm orb}^{\rm i}$)=(1.26562$\,\rm M_{\odot}$, 0.2$\,\rm M_{\odot}$, 0.045\,d) (see Set.\,5 in Tab.\,1). The purple and blue lines represent evolutionary stages before and after the AIC process, respectively.}
  \end{center}
\end{figure}

\subsection{4U 1626$-$67}
4U 1626$-$67 is one of the newly formed NSs companying with an ultra-light star, which could be a possible evidence for the existence of the AIC events (e.g. Tauris et al. 2013).
According to the relationship between the energy of the fundamental cyclotron harmonic and the NS magnetic field strength, Orlandini et al. (1998) estimated that the magnetic field for the pulsar in 4U 1626$-$67 is about $3\times 10^{\rm 12}\,\rm G$, which indicates that the NS almost have not accrete any material after its formation; whereas the orbital period is 0.028 d and the companion mass is about 0.01$-$0.02$\,\rm M_{\odot}$, which implies that the companion has transferred a considerable amount of material onto the primary (Verbunt et al. 1990; Yungelson et al. 2002).
If it does originates from the AIC channel, 4U 1626$-$67 corresponds to the phase when the companion just refills its Roche-lobe after the AIC process due to the detection of X-ray irradiation that is caused by the mass-accretion of NS.
Additionally, the theoretical companion mass and the orbital period at this phase should be comparable with the speculated values.

Fig.\,4 shows a possible evolutionary path (see set 5 in Tab.\,1) for the formation of 4U 1626$-$67.
The initial binary consists of a 1.26562$\,\rm M_{\odot}$ ONe WD and a 0.2$\,\rm M_{\odot}$ He WD with an orbital period of 0.045 d.
The binary evolution is similar to the example shown in Fig.\,1.
After about 400 Myr mass-transfer process, the ONe WD increases its mass to the Chandrasekhar mass limit and an AIC event occurs.
After that, the orbital period is consistent with the observed value of 4U 1626$-$67 ($\sim$0.028 d) when the He WD refills its Roche-lobe.
Due to the mass-radius relationship of He WDs, the mass of the lobe filling He WDs should be similar for a given accretor mass and orbital period.
In this case, we can speculate that the mass of the donor in 4U 1626$-$67 is about 0.0127$\,\rm M_{\odot}$, which is consistent with the speculated mass (0.01$-$0.02$\,\rm M_{\odot}$) in previous studies (Verbunt et al. 1990; Yungelson 2002).
During the following 1.15 Gyr, the binary behaves as a UCXB.
Finally, the NS will increase its mass to about 1.2522$\,\rm M_{\odot}$ and the companion mass is about 0.0084$\,\rm M_{\odot}$ with an orbital period of 0.038 d.

It is notable that there are some disadvantages for the ONe WD$+$He WD channel to explain the formation of 4U 1626$-$67. 
(1) If 4U 1626$-$67 is a semidetached ONe WD+He WD system, the He lines in its spectrum should be easily detectable (see, e.g., Werner, K. et al. 1984).
However, the double-peaked O and Ne emission lines in its X-ray spectrum and the strong C and O lines in its UV spectrum without He lines indicate that the companion in 4U 1626-67 may be an evolved He star or even a hybrid WD (Nelemans et al. 2010), which is contrary to the assumption that the companion is a He WD.
(2) The theoretical mass-transfer rate is about $10^{-11}\,\rm M_{\odot}yr^{-1}$ when the He WD refills its Roche-lobe, which is lower than the observed rate in 4U 1626-67 estimated based on the observed spin-up rate and simple angular momentum conservation considerations (Yungelson 2002; see Fig.\,3). 
This deficiency has already been claimed by previous theoretical studies (e.g. Yungelson 2002).
A possible explanation is that the observable mass-transfer rate speculated by the detected spin-up of the NS corresponds to the accretion rate from an unstable accretion disc, whereas the actual mass-transfer rate may be consistent with theoretical models (Schulz et al. 2001; Heinke et al. 2013). 
(3) It has been suggested that the AIC events would produce NSs with fast spin (e.g. Phinney \& Kulkarni 1994), which is contact with the long spin period of 7680 ms for the pulsar in 4U 1626-67. 
For more discussions about the spin of NSs from AIC process see Sect.\,4.

\subsection{Gravitational wave sources}
\begin{figure}
\begin{center}
\epsfig{file=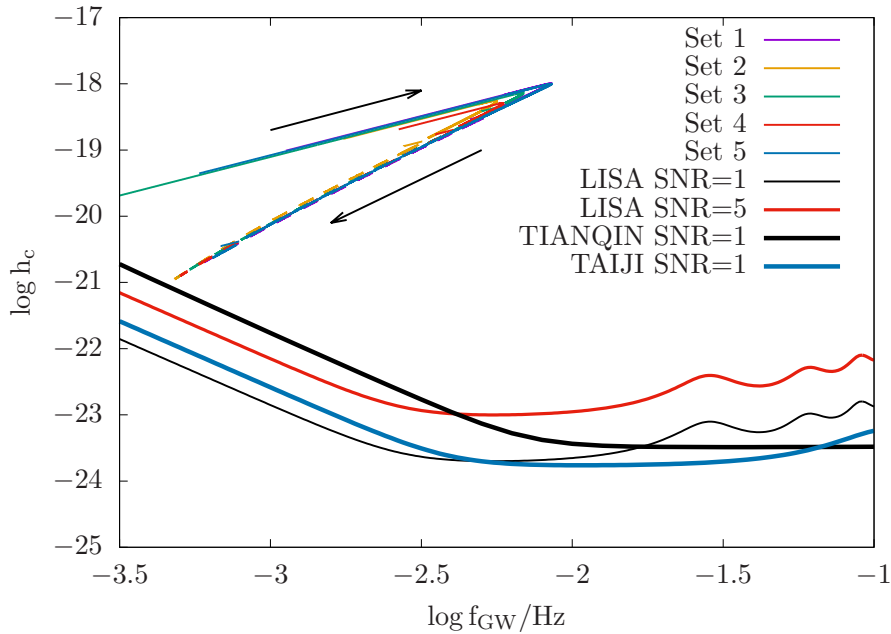,angle=0,width=12.5cm}
 \caption{Evolutionary tracks of five typical UCXBs in the characteristic strain amplitude$-$the GW frequency ($\log f_{\rm GW}$$-$$\log h_{\rm c}$) diagram, in which the distance from the sources to the detectors is assumed to be 15 kpc.
The solid and dotted lines present evolutionary phases before and after the AIC process, respectively.
We also plot the sensitivity curve of the future space-based gravitational wave telescope LISA (Robson, Cornish \& Liu 2019), Taiji (Ruan et al. 2020) and TianQin (Wang et al. 2019).}
  \end{center}
\end{figure}

Recently, the gravitational wave signals from the double black hole merger (e.g. GW150914) and double NSs (e.g. GW170817) are successfully detected by the ground$-$based Advanced Laser Interferometer Gravitational-Wave Observatory (see Abbott et al. 2016, 2017).
Nowadays, many space$-$based gravitational wave telescopes, like LISA, Taiji and TianQin are on their way (e.g. Danzmann \& LISA Study Team 1997; Luo et al. 2016; Ruan et al. 2020).
Compared with the ground$-$based gravitational wave telescope, the space$-$based ones can detect gravitational wave with lower frequency.
In this case, the gravitational wave signals from double WDs would be detectable (e.g. Liu et al. 2020).

In this work, the prescription for calculating the characteristic strain amplitude and the chirp mass are similar to those described in Wang et al. (2021). The distance is supposed to be 15 kpc.
Fig.\,5 presents the evolutionary tracks of the selected binaries (see Tab.\,1) in the gravitational wave frequency$-$the characteristic  strain amplitude ($\log f_{\rm GW}$$-$$\log h_{\rm c}$) plane.
From this figure, we can see that the the gravitational wave frequency is lower than $\sim$0.01 Hz, and the characteristic strain amplitude is in the range of $\sim$$10^{\rm -21}$ to $\sim$$10^{\rm -18}$.
This figure indicates that the semidetached ONe WD$+$He WD systems and the following formed UCXBs are both detectable for the future space based gravitational wave telescopes.

\section{Discussion and summary} \label{4. Discussion}

The ONe WD$+$He WD systems will evolve to NS$+$He WD systems after the AIC events, during which the binaries will act as UCXBs when the He WDs fill their Roche-lobes again.
Wang et al. (2021) suggested that the relation between the mass-transfer rate on the declining stage and the orbital period can be used to distinguish the mass donors in UCXBs.
As shown in Fig.\,3, the UCXBs from the He WD donor channel of AIC events locate between the evolutionary tracks of the NS$+$He WD channel and NS$+$CO WD channel. 
Fig.\,3 shows that the UCXB XTE J1751$-$305 can only be covered by the He WD donor channel.
Moreover, recent asteroseismology model indicates that the accreting material onto the NS in XTE J1751$-$305 composed mostly of He (see Lee 2014; Strohmayer \& Mahmoodifar 2014).
Thus, we can speculate that XTE J1751$-$305 may originate from an AIC event with a He WD donor.

Basically, ONe WD+He star systems are thought to evolve to intermediate-mass binary pulsars will short orbital periods if the ONe WDs can grow in mass approach to the Chandrasekhar mass limit (e.g., Liu et al. 2018; Ablimit 2022).
In these binaries, the He stars would form CO/ONe cores before or during the mass-transfer process.
In the present work, we found that some ONe WD+He star systems can also evolve to UCXBs after the AIC processes, in which the He stars are less massive and the orbital periods are shorter compared with the ONe WD+He star systems in Liu et al. (2018). 
For example, a 1.25$\,\rm M_{\odot}$ ONe WD accompanying by a 0.32$\,\rm M_{\odot}$ He star with an orbital period of 0.025 d could produce an AIC event when the donor mass is 0.135$\,\rm M_{\odot}$ and the orbital period is 0.0088 d.
When the donor star refills its Roche-lobe after the AIC process, the system would behave as an UCXB during the following 8 Gyr.
Finally, the binary evolves to a 1.3151$\,\rm M_{\odot}$ NS and a 0.0053$\,\rm M_{\odot}$ donor with an orbital period of 0.0516 d when the code stops. 

The present work studied the formation of AIC events with He WD donors, and explored the properties of the formed NS$+$He WD systems just after the AIC events.
We suggest that this channel is a possible path to explain the formation of a portion of the newly formed NSs with ultra-light companions. However, the magnetic field and spin properties of NSs originated from the AIC process are still quite unclear, which is quite dependent on the magnetic field and spin properties of its progenitor WD. 
It has been suggested that about 15 to 20\% WDs have magnetic field in the order of $10^{\rm 3}\,\rm G$ (e.g. Jordan et al. 2007).
For a Chandrasekhar mass WD with a magnetic field of $10^{\rm 3}\,\rm G$, it seems that the magnetic field of the newly formed NS would be as high as $10^{\rm 8}\,\rm G$, and the spin period would as low as a few millisecond even just a small fraction of the WD angular momentum is considered to be conserved (Dessart et al. 2006).
However, as discussed in Tauris et al. (2013), the magnetic field of the millisecond pulsars are typically 5 magnitude lower than that in young normal pulsars, and the observed properties of almost all of the known young NSs that are associated with supernova remnant are inconsistent with the typical millisecond pulsars, i.e. fast spin and low magnetic field. Further studies on the properties of the NSs from the AIC events are needed.

Nelson et al. (1986) studied the formation of ultrashort period binaries including 4U 1626$-$67 by evolving some compact star$+$H-depleted star binaries, in which the secondaries are treated as a composite polytrope.
Yungelson et al. (2002) suggested that 4U 1626$-$67 may originate from the AIC channel with a hybrid HeCO WD donor (0.3$-$0.4$\,\rm M_{\odot}$).
The main difference between Yungelson et al. (2002) and the present work is the different composition in the donors and thus different accumulation rate for the accreted material on the surface of the ONe WDs.
Additionally, there are also some other detected strong magnetic field NSs with ultra-light companions, e.g. GRO J1744-28, PSR J1744-3922 and PSR B1831-00 with significant longer orbital period (11.8 d, 0.2 d and 1.8 d, respectively).
There may be some physical processes that are not considered in the present work or other models to explain their formation, which still need further investigations.

In the present work, we investigate the evolution of a series of ONe WD$+$He WD systems to the formation of NSs with ultra-light companions through the AIC processes, and then explore the properties of the UCXBs during the recycling process.
By using the He WD donor channel of the AIC events, we reproduced the detected companion mass and orbital period of 4U 1626-67, which is one of the newly formed NSs with ultra-light companions detected in the Galactic disc.
Combined with previous asteroseismology results, we speculated that the UCXB XTE J1751$-$305 originate from the AIC events  with He WD donors.
We also found that the gravitational signals of the merging process before and after the AIC process are detectable for future space-based gravitational wave telescopes.

\section{DATA AVAILABILITY}
The data and code of the numerical simulation in this work are available on request by contacting DDL.

\section*{Acknowledgments}
We acknowledge useful comments and suggestions from the anonymous referee.
The present work is supported by the National Natural Science Foundation of China (Nos 12273105 and 12225304), the National Key R\&D Program of China (No. 2021YFA1600403/04), the Youth Innovation Promotion Association CAS (No. 2021058), the Western Light Project of CAS (No. XBZG-ZDSYS-202117), the science research grants from the China Manned Space Project (Nos CMS-CSST-2021-A12/B07) and the Yunnan Fundamental Research Projects (Nos 202001AU070054, 202101AT070027, 202101AW070047, 202001AS070029 and 202201BC070003), and the Frontier Scientific Research Program of Deep Space Exploration Laboratory (No. 2022-QYKYJH-ZYTS-016).

\label{lastpage}
\end{document}